\documentclass[journal=nanolett,manuscript=article]{achemso}

\usepackage{amsmath}   
\usepackage{amssymb}
\usepackage{braket}
\usepackage{leftidx}
\usepackage{graphicx}    
\usepackage{verbatim}   
\usepackage{color}         
\usepackage{subfigure}  
\usepackage{dcolumn}    
\usepackage{textcomp}
\usepackage{lipsum}
\usepackage{threeparttable}
\usepackage{multirow}
\usepackage{siunitx}
\usepackage{tabularx}
\usepackage{adjustbox}
\setkeys{acs}{email=false}

\usepackage{graphicx,amsmath,amssymb,bm}
\usepackage{hyperref,verbatim}
\usepackage{bbold}
\hypersetup{%
  breaklinks = {true},
  citecolor = {blue},
  colorlinks = {true},
  linkcolor = {red},
  pdfcreator = {\LaTeX\ and \flqq hyperref\frqq},
  pdffitwindow = {true},
  pdfmenubar = {true},
  pdfpagelayout = {SinglePage},
  pdfstartview = {Fit},
  pdftoolbar = {true},
  plainpages = {false},
}

\usepackage{siunitx}

\renewcommand{\eqref}[1]{Eq.~(\ref{#1})}

\RequirePackage[usenames,dvipsnames]{xcolor}
\usepackage{soul}

\newcommand{\cea}{Univ. Grenoble Alpes, CEA, CNRS, Spintec, 38000 Grenoble, France}
\newcommand{\nimte}{Key Laboratory of Magnetic Materials and Devices, Ningbo Institute of Materials Technology and Engineering, Chinese Academy of Sciences, Ningbo, China}
\newcommand{\Thales}{Unité Mixte de Physique, CNRS, Thales, Université Paris-Saclay, 91767 Palaiseau, France}
\newcommand{\iuf}{Institut Universitaire de France, 75231, Paris, France}

\author{Ali Hallal}  \affiliation{\cea}%
\author{Jinghua Liang}   \affiliation{\nimte}%
\author{Fatima Ibrahim} \affiliation{\cea}%
\author{Hongxin Yang} \affiliation{\nimte}%
\author{Albert Fert} \affiliation{\Thales}%
\author{Mairbek Chshiev} \affiliation{\cea} \altaffiliation{\iuf}%

\title{Rashba-type Dzyaloshinskii-Moriya interaction, perpendicular magnetic anisotropy and skyrmion states at 2D materials/Co interfaces}

\begin{document}


\begin{abstract}
We report a significant Dzyaloshinskii-Moriya interaction (DMI) and perpendicular magnetic anisotropy (PMA) at interfaces comprising hexagonal boron nitride (h-BN) and Co. By comparing the behavior of these phenomena at graphene/Co and h-BN/Co interfaces, it is found that the DMI in latter increases as a function of Co thickness and beyond three monolayers stabilizes with one order of magnitude larger values compared to those at graphene/Co, where the DMI shows opposite decreasing behavior. At the same time, the PMA for both systems shows similar trends with larger values for graphene/Co and no significant variations for all thickness ranges of Co. Furthermore, using micromagnetic simulations we demonstrate that such significant DMI and PMA values remaining stable over large range of Co thickness give rise to formation of skyrmions with small applied external fields in the range of 200-250 mT up to 100 K temperatures. These findings open up further possibilities towards integrating two-dimensional (2D) materials in spin-orbitronics devices.
\end{abstract}


\maketitle


For years, interfaces comprising magnetic transition metal (e.g. Co or Fe), oxides (e.g. MgO) or non-magnetic metals (e.g. Pt) have been of central interest due to a large variety of spintronics phenomena enabling new generations of applications~\cite{RMP1,RMP2}. Two dimensional (2D) materials such as graphene and hexagonal boron nitride (h-BN), are appealing as novel materials with exceptional properties~\cite{Geim2007, Zhang2005, Yazyev2008, Yazyev2008PRL, wang2009, Tombros2008, Shahil2012, Khan2012, Yu2007, Huang2011, Liu2012} that can replace conventional ones and open new prospects for information technology. Graphene has been widely studied for use in both lateral and vertical spintronic devices. Various approaches have been proposed to inject, detect, and induce spin polarized currents in graphene and other 2D materials~\cite{Han2014, Muhammad2018, stephan2015, Yazyev2008, Lazi2014, Maassen2011, Yang2013, Hallal2017}. Due to their protective nature against oxidation, they have been incorporated in magnetic tunnel junctions catalyzing at the same time spin-dependent transport and magnetic properties of these structures~\cite{piquemal2018}. Furthermore, despite its weak spin-orbit coupling (SOC), graphene coating of Co is found to induce a large perpendicular magnetic anisotropy (PMA)~\cite{Yang2016} and a significant Dzyaloshinskii-Moriya interaction (DMI) due to Rashba effect~\cite{Yang2018}. Those findings have promoted graphene, and subsequently 2D materials, as potential candidates for novel low-power spin-orbitronic based applications.

In this letter, using first principles calculations, we demonstrate a significant PMA and a large DMI at Co/2D materials (Gr or h-BN) interfaces. We find that PMA at the Co/h-BN interface is preserved as in the case of graphene coverage, thanks to the hybridization between $d{_Z{^2}}$ orbitals of Co with $P_Z$ of 2D materials. Furthermore, we demonstrate that the DMI coefficient at Co/h-BN interface increases as a function of Co thickness up to 3ML reaching its maximum value of 1.2 meV beyond which it saturates for higher thicknesses. This tendency is found to be opposite to that at Co/Gr interface where DMI decreases with the Co thickness. The Rashba splitting in case of Co/h-BN is found to be at least twice larger than that of Co/Gr giving rise to the larger DMI values. The difference between the two interfaces is explained by the presence of two competing dipoles at the Co/Gr interface compared to only one dipole at the opposite Co/h-BN interface which gives rise to a larger DMI. Moreover, our micromagnetic simulations show the possibility of skyrmions formation at h-BN(AC)/Co interface with the application of small external magnetic field and remains stable up to hundred Kelvin.

\begin{figure*}[htp]
	\centering
	\includegraphics[clip = true, trim=0cm 5.2cm 0cm 4.5cm, width=1.0\textwidth]{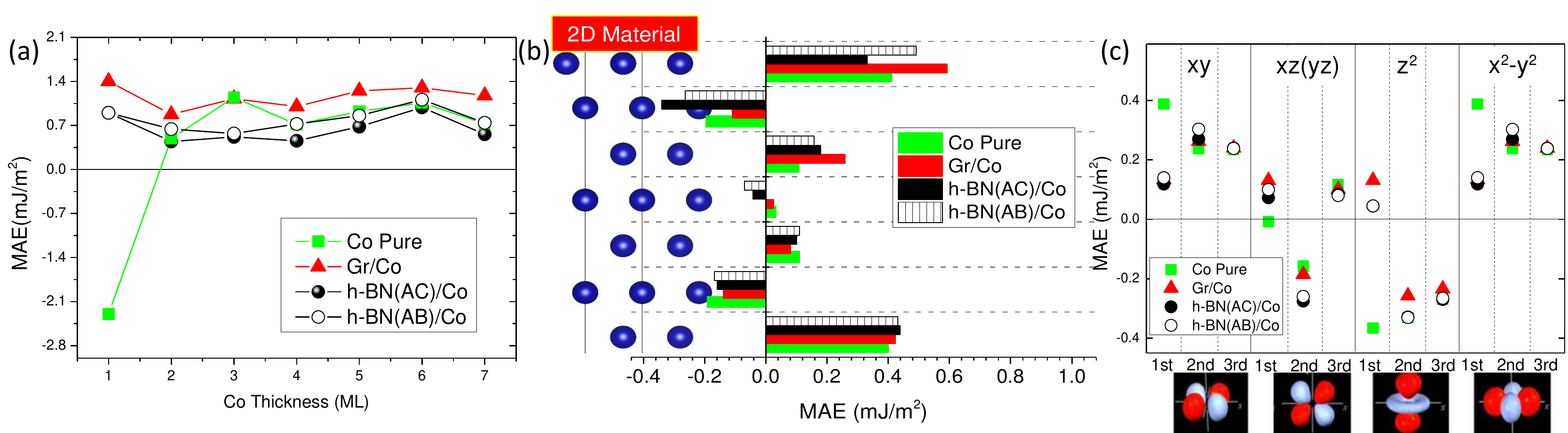}
	\caption{(a) Magnetic anisotropy energy  of Co slab, Gr/Co, h-BN(AB)/Co and h-BN(AC) interfaces as function of Co thickness, (b) Layer resolved contributuions into MAE for Co slab and 2D Material/Co interfaces comprising 7 ML of Co. 2D material on top of Co is represented schematically by red box, Co atoms are represented by blue balls. (c) d orbital resolved contributions into MAE from the 1st, 2nd and 3rd Co monolayers adjacent to the 2D material. MAE values in all panels for Co slab, Gr/Co, h-BN(AB)/Co and h-BN(AC) are indicated by green squares, red triangles, solid black and open white circles, respectively. The same color code apply to horizontal bars representing MAE contributions in (b).	}
	\label{1}
\end{figure*}

The magnetic anisotropy energy (MAE) is calculated by taking the difference in total energy for in-plane ($E_\parallel$) and out-of plane ($E_\perp$) magnetization orientations:
\begin{equation*}
MAE= \frac{E_\parallel - E_\perp}{ A}
\label{eq:1b}
\end{equation*}
where $A$ represents the area of hexagonal structure and it is equal to $A=a^2 \frac{\sqrt{3}}{2}$. The MAE behavior for Gr/Co, h-BN(AC)/Co, and h-BN(AB)/Co interfaces as a function of Co thickness are presented in Fig.~\ref{1}(a) by red triangles, black and white circles, respectively. To highlight the effect of 2D materials on MAE, we show for comparison the magnetic anisotropy of a bare Co slab shown by green squares in Fig.~\ref{1}(a). For a single layer of Co, the magnetization strongly favors the in-plane direction. One can see that the situation changes when Co is coated by single layer of 2D materials which induce the PMA by adding ~3.2 or ~3.7~mJ/m$^2$ to MAE in case of h-BN and Gr coverage, respectively. For thicker hcp Co(0001) films, perpendicular magnetic anisotropy is maintained in all cases considered including the one with graphene coating which further enhances the PMA in agreement with previous reports~\cite{Yang2016}.  

To further compare and understand mechanisms of the catalyzing effect of h-BN and graphene on PMA in Co, we present the layer resolved MAE contributions for different structures with 7 ML of Co as shown in Fig.~\ref{1}(b). Overall, MAE shows a clear attenuated oscillatory behavior as a function of the distance from  interface. Such oscillatory behavior has been also reported in different systems such as transition metal/oxide interfaces~\cite{Hallal2013}. In general, the main contribution to the PMA is located at the first interfacial Co layers with some significant contribution arising from the third monolayer as well. Since magnetic anisotropy is a surface phenomenon, the effect of different coatings impact mostly the first three atomic layers as shown in Fig.~\ref{1}(b). Graphene coating further enhances the PMA contribution of the first and third mono-layers and reduces the in-plane magnetic anisotropy (IMA) of the second layer~\cite{Yang2016}. In case of h-BN coating, the PMA contribution from the first atomic layer depends on the type of stacking: while h-BN(AC) coating shows a PMA decrease, h-BN(AB)/Co leads to the PMA enhancement. However, in both cases  MAE contribution of the 2nd and 3rd ML favoring IMA and PMA, respectively, are further enhanced compared to those of bare Co slab. To elucidate the microscopic origin of PMA, Fig.~\ref{1}(c) shows the orbital resolved contributions of the first, second, and third Co atomic layers in 2D/Co structures with 5 ML of Co to simplify presentation. One can see, that in the case of bare Co, the PMA contribution from the first layer originates from in-plane orbital hybridizations, namely d$_{xy}$ and d$_{x^2-y^2}$. On the other hand, the ones with out-of-plane orbital d$_{z^2}$ favors strong IMA while those with  d$_{xz(yz)}$ orbitals have a negligible MAE contributions.
Graphene and h-BN coating decreases the PMA contribution of in-plane orbitals hybridization of the first layer. However, due to the strong hybridization between d$_z^2$ of Co with P$_z$ of 2D materials the out-of-plane orbitals favor PMA.

\begin{figure*}[htp]
	\centering
	\includegraphics[clip = true, trim=6cm 0.1cm 6cm 0.1cm, width=1.0\textwidth]{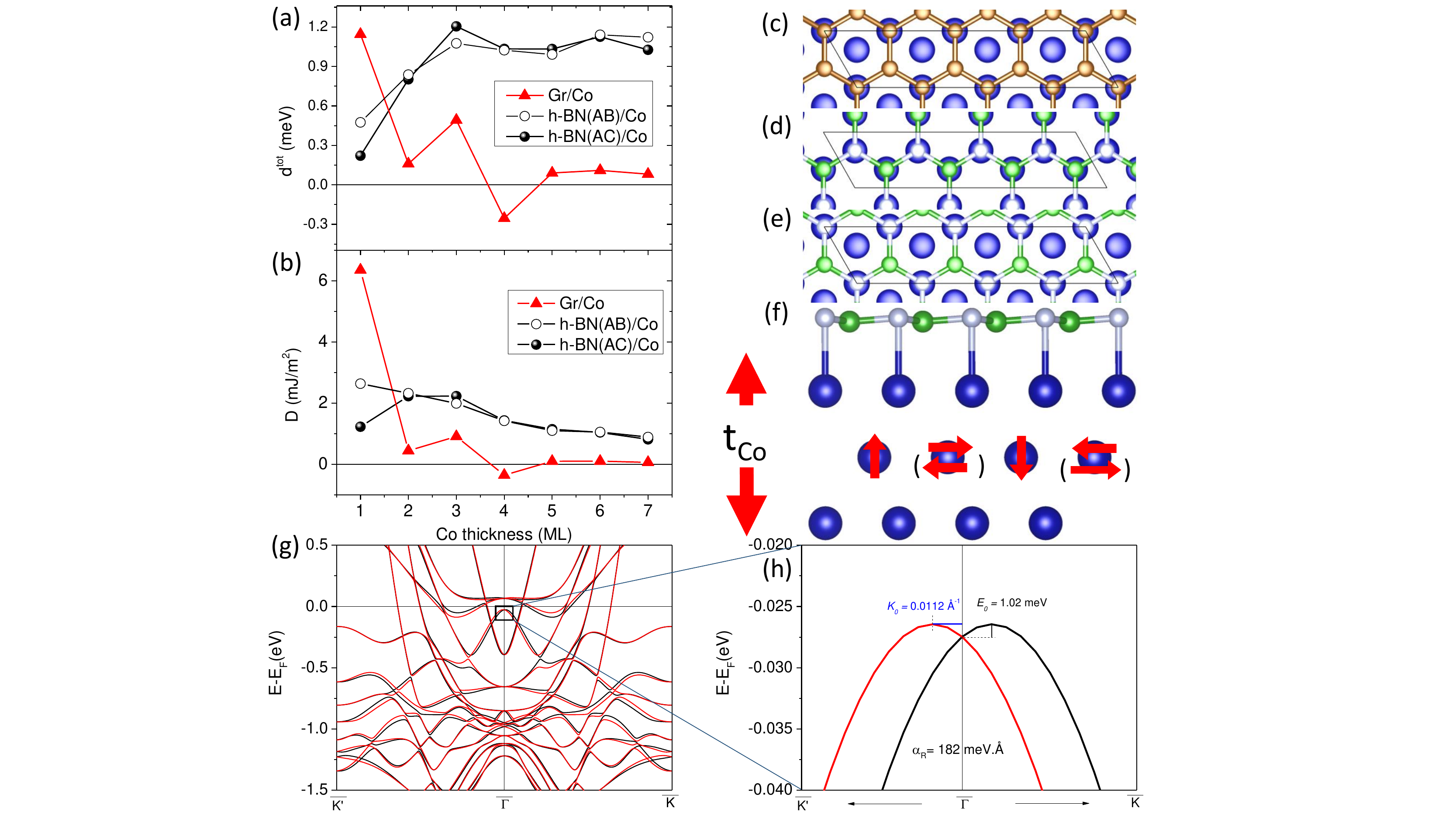}
	\caption{(a) and (b) Microscopic and micromagnetic DMI coefficients of Gr/Co, h-BN(AB)/Co and h-BN(AC)/Co interfaces as a function of Co thickness. (c),(d) and (e) Top views of Gr/Co, h-BN(AB)/Co and h-BN(AC)/Co structures, respectively. (f) Side view of Gr/Co(0001) with schematic representation of clockwise (anti-clockwise) spin configurations indicated by arrows. (g) and (h) Band structures for h-BN(AC)/Co slab with magnetization axis along $ \left<1 1 \bar{2}0\right >  $ (black) and  $ <\bar{1} \bar{1} 20> $ (red) used to estimate the Rashba splitting.  Carbon, Boron, Nitrogen and Cobalt atoms are represented by brown, green, grey, and blue balls, respectively. }
	\label{3}
\end{figure*}

Next, we present the results of DMI calculations for these 2D materials/Co interfaces. We employed the approach based on chirality-dependent total energy calculations applied previously for Pt/Co, MgO/Co and Gr/Co structures to calculate the microscopic and micromagnetic DMI constants, $d^{tot}$ and $D$, respectively~\cite{Yang2015,Yang2018sci,Yang2018}. In Fig.~\ref{3}(a) and (b) we show the dependence of DMI coefficients  as a function of Co thickness up to 7 ML for Gr/Co, h-BN(AB)/Co and h-BN(AC)/Co interfaces shown in Fig.~\ref{3}(c),(d) and (e), respectively. The arrows in Fig~\ref{3}(f) schematically indicate spin configuration with clockwise (anticlockwise) chirality. The microscopic constant $d^{tot}$ at Gr/Co interface exhibits a large DMI value of 1.2~meV for 1ML of Co, in agreement with previous reports~\cite{Yang2018}. As Co thickness increases, the DMI strength decreases and even changes a sign at 4ML thickness but restores left-handed chirality above 5ML of Co thickness saturating at ~0.15~meV. Interestingly, h-BN/Co interface shows an opposite DMI trend with Co thickness.  For 1ML of Co, the $d^{tot}$ of h-BN/Co is about 0.2~meV and 0.5~meV for AC and AB stacking, respectively. With the increase of Co thickness, the microscopic DMI increases as well up to 3 MLs and then saturates around 1.1~meV for both stackings. Thus, the DMI strength at h-BN/Co interface with large Co thickness is one order of magnitude larger than that at graphene/Co and found to be robust against the stacking order. Despite the fact that largest value of the micromagnetic DMI, $D=6.5$~mJ/m$^2$, is found for 1~ML of Co coated with graphene, in this case the DMI weakens with increasing the Co thickness becoming negligible for Co films avove 5 ML. For thin Co films, the tendency of micromagnetic DMI constant behavior at small Co thicknesses depends on the stacking order of h-BN/Co interface. For AB stacking, we observe a monotonic behavior where $D$ decreases as a function of Co thickness[Fig~\ref{3}(b)]. However, D exhibits a non-monotonic behavior in the case of AC stacking: it increases for Co thicknesses up to 3ML and decreases afterwards. Interestingly, the micromagnetic DMI for thick Co layers is independent of the stacking order with significantly larger values compared to Gr/Co interface for both AB and AC stackings. These results suggest that h-BN coating provides strong and robust DMI for thick Co films which combined with aforementioned PMA findings represent major interest for 2D spin-orbitronics including skyrmionics~\cite{Sampaio2013, Gibertini2019}.

\begin{figure}[htp]
	\centering
	\includegraphics[clip = true, trim=5cm 2cm 7cm 2cm, width=0.5\textwidth]{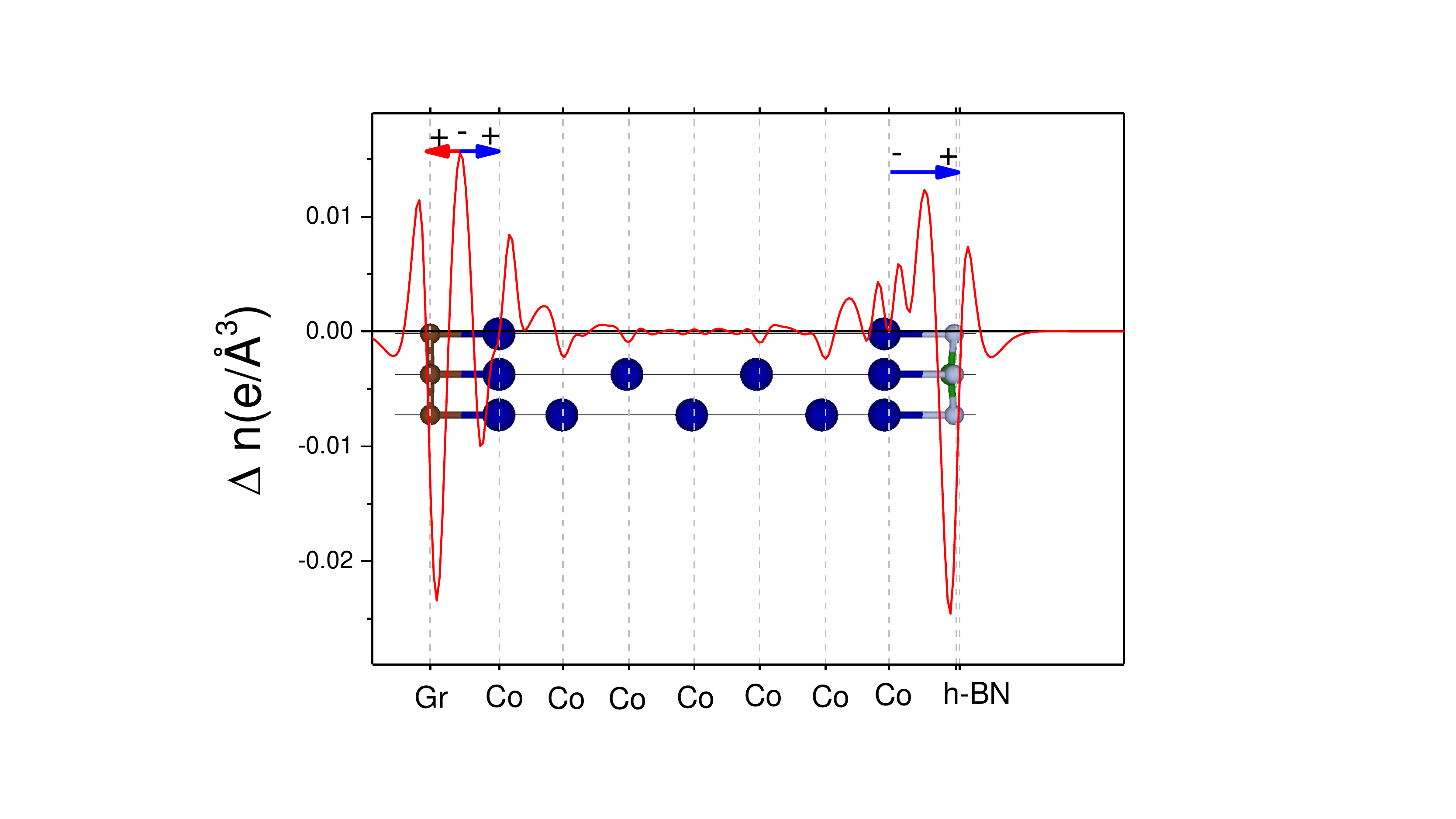}
	\caption{Planar charge difference for Gr/Co/h-BN heterostructure shows the presence of two competing dipoles at the Gr/Co interface unlike Co/h-BN one schematically shown within the figure. Cobalt, Carbon, Boron and Nitrogen atoms are represented by blue, brown, grey and green balls, respectively. }
	\label{4}
\end{figure}

The physical mechanism behind DMI in conventional metallic interfaces such as Co/Pt is governed by Fert-Levy model~\cite{Yang2015}. In the absence of heavy atoms, as in the cases we address here, the DMI mechanism is dominated by Rashba-type model~\cite{Yang2018}. The micromagnetic DMI can be roughly expressed as $D = 2k_{R} A_{ex}$~\cite{Kim2013}, where $A_{ex}$ indicates the interfacial exchange stiffness and $k_{R} = \frac{2\alpha{_R} m_{e}}{\hbar^{2}} $ is determined by the Rashba coefficient $\alpha_{R}$, effective electron mass $m_e$ with $\hbar$ being the reduced Planck constant. The effective mass of Co is taken to be 0.45 $m_0$ (with $m_0$ being the electron rest mass), and the interfacial exchange stiffness $A$ is about 5 pJm$^{-1}$ for Co thin films. For Gr/Co interface the Rashba splitting is found to be about 82 meV.\AA{} as reported in Ref.~\cite{Yang2018}. Following the same method, we estimated the Rashba splitting at h-BN(AB)/Co and h-BN(AC)/Co interfaces to be 225 and 182 meV.\AA{}, respectively [see Fig.~\ref{3}(g) and (h)]. The micromagnetic DMI estimated from these Rashba splittings at Gr/Co, h-BN(AB)/Co, and h-BN(AC)/Co are found to be ~0.98~mJ/m$^2$, 2.66~mJ/m$^2$ and 2.15 mJ/m$^2$ for structures with 3 ML thick Co, respectively. The DMI values obtained from first principles calculations and estimated from the Rashba splittings are listed in Table~\ref{T1}. Using micromagnetic DMI values, we calculated the microscopic ones to be ~0.52~meV, 1.44~meV and 1.16~meV for graphene, h-BN(AB), and h-BN(AC) coatings, respectively. Those values are in good agreement with the DMI ones obtained from first principles calculations. One should note that the estimated Rashba splitting at h-BN/Co interface is comparable to that at MgO/Co reported in Ref.~\cite{Yang2018sci}. However, the DMI value calculated using first principles approaches is almost twice larger for MgO/Co interface. This difference could be attributed to the different $A_{ex}$ values in different systems. In Table~\ref{T1}, we also list experimental values of the microscopic DMI at Gr/Co/Ru~\cite{Yang2018} and  Gr/Co/Pt~\cite{Ajejas2018} systems for comparison. The extracted values of Gr/Co from aforementioned works show opposite signs with 0.16~meV and -0.6~meV for structures on Ru and Pt substrate, respectively. This difference between both experiments is most probably due to different Co thickness used. For Pt substrate, the Co thickness is about 4~ML which is in line with our first principles calculations showing that for this particular thickness the DMI value is negative. However, the DMI becomes positive again for larger Co thickness and is about 0.15~meV in perfect agreement with DMI obtained with Ru substrate.

\begin{table}[htp]
	\centering
	\renewcommand{\arraystretch}{1.2}
	\begin{adjustbox}{max width=\textwidth}
		\begin{tabular}{|c|c|c|c|c|c|c|c|c|c|c|c|c|}
			\hline
			\multirow{2}{2cm}{   } & \multicolumn{6}{c|}{\textbf{$d^{tot}$(meV)}} & \multicolumn{2}{c|}{\textbf{\textbf{$d^{tot}$(meV)}}} & \multicolumn{3}{c|}{\textbf{\textbf { DMI  }}}     \\
			\multirow{2}{2cm}{   } & \multicolumn{6}{c|}{\textbf{This work}} & \multicolumn{2}{c|}{\textbf{ Experiments }} &  \multicolumn{3}{c|}{\textbf{\textbf { from Rashba }}}  \\
			\cline{1-12}
			&     &     &     &     &     &     & Gr/Co/Ru~\cite{Yang2018}  & Gr/Co/Pt~\cite{Ajejas2018}&  {$\alpha$(meV.\AA{})}  & D (mJ/m$^2$)   &  d (meV) \\
			Co Thickness & 1ML & 2ML & 3ML & 4ML & 5ML & 6ML &      4-6ML           &  4ML &  &  3ML & 3ML \\
			\hline

			Gr/Co   		& 1.14 & 0.16 & 0.49 & -0.24 & 0.09 & 0.11 & 0.16$\pm$0.05 & -0.6$\pm$0.15   & 82~\cite{Yang2018} &   0.98   &   0.52   \\ \hline
			h-BN(AB)/Co  & 0.47 & 0.84 & 1.07 &  1.02 & 0.99 & 1.14 &               &                 &    225        &   2.66   &   1.44   \\ \hline
			h-BN(AC)/Co  & 0.22 & 0.80 & 1.21 &  1.03 & 1.03 & 1.13 &               &                 &    182        &   2.15   &   1.16   \\ \hline   
			Co/Ru       &      & 	 &   	&       &      &      & 	-0.05	  &                 &               &          &          \\ \hline
			Co/Pt       &      &      &      & 	    &      &      &               &     1.05        &               &          &          \\ \hline
			Gr/Co/Ru(Pt) &      &      &      &       &      &      &     0.11      &   0.45$\pm$0.15 &               &          &          \\ \hline
			MgO/Co~\cite{Yang2018sci} &      &      &   2  &       &      &      &               &                 &     224       &   2.65   &   1.43   \\ \hline
		\end{tabular}
	\end{adjustbox}
	\caption{A comparison between the calculated DMI coefficients and the experimental values obtained from Ref.~\cite{Yang2018,Ajejas2018} with different substrates and Co thickness for Gr/Co interfaces, the DMI coefficients and estimated Rashba splitting for h-bn(AB)/Co and h-BN(AC)/Co interfaces are also listed.}
	\label{T1}
\end{table}

In order to understand the difference between microscopic mechanisms of Gr/Co and h-BN/Co interfaces, in Fig.~\ref{4} we show the induced charge densities across Gr/Co/h-BN heterostructure.  One can note two competing opposite electric dipoles present at the Gr/Co interface, while at the h-BN/Co one there exists only one dipole pointing away from the Co film. This larger interfacial dipole is at the origin of the larger Rashba splitting and consequenlty stronger DMI at the h-BN/Co interface compared to the Gr/Co one.

\begin{figure}[tp]
\centering
\includegraphics[clip = true, trim= 0.6cm 4.5cm 2.0cm 2.0cm, width=1\textwidth]{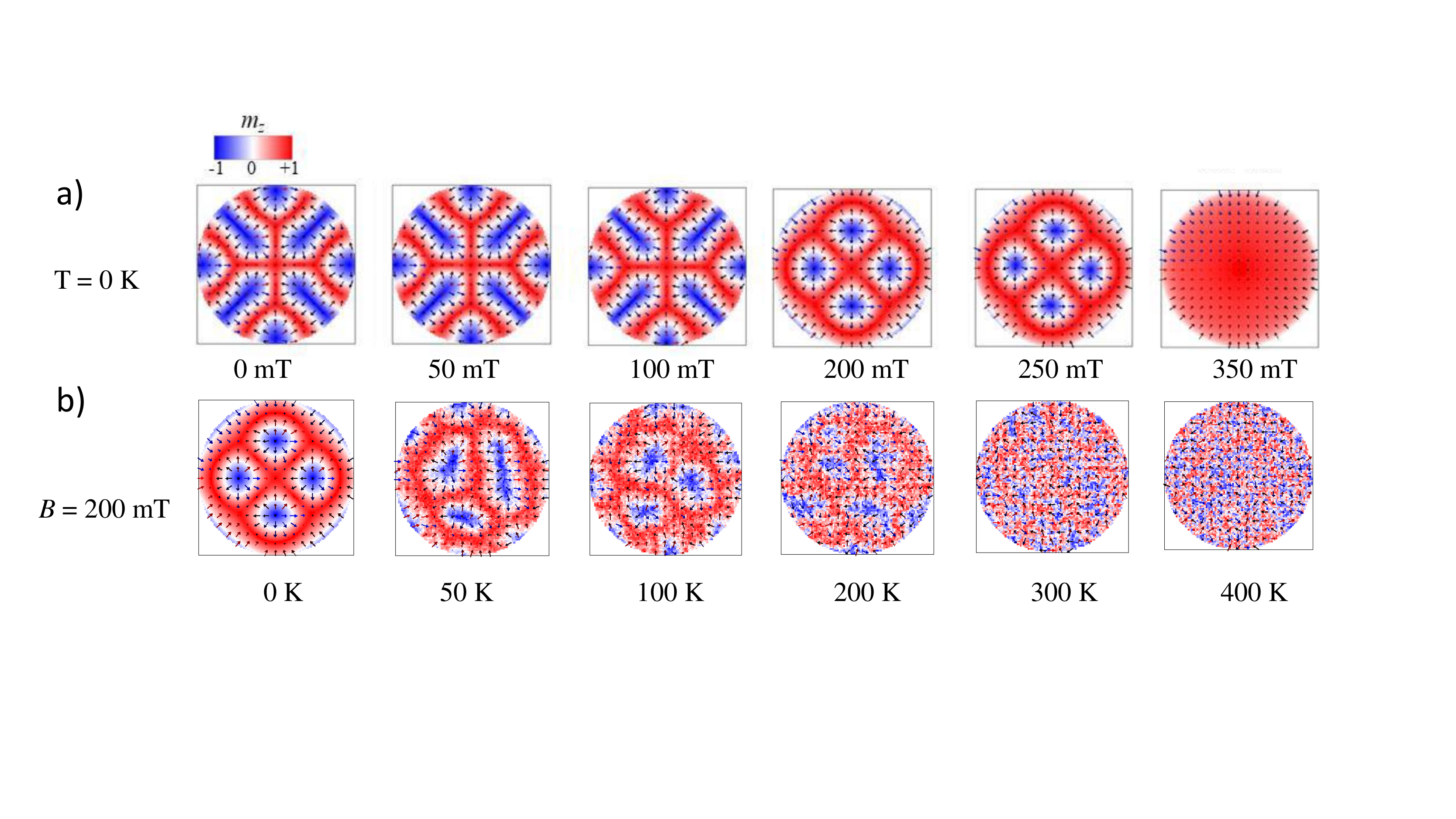}
\caption{The relaxed magnetization distributions of 100-nm-wide h-BN(AC)/Co(3 ML) a) at different external magnetic field  for  $0 K$ Temperature  b)  at finite Temperatures for 200mT external magnetic field. The initial state is a uniform out-of-plane ferromagnetic state. The in-plane and out-of-plane components are indicated by the arrows and the color map, respectively.
}
\label{5}
\end{figure}

Finally, with significant PMA and DMI values at the h-BN/Co interfaces, it is interesting to explore a possibility to induce and stabilize chiral spin textures including skyrmions in such heterostructures. We performed micromagnetic simulations using the Object Oriented MicroMagnetic Framework (OOMMF)~\cite{Donahue} that solves the spin dynamics based on the Landau-Lifshitz-Gilbert equation~\cite{Gilbert}. As an example, we considered the h-BN(AC)/Co(3 ML) heterostructure, which has the largest microscopic DMI among the investigated structures, with corresponding magnetic material parameters for exchange $A$ = 5~pJ/m, DMI $D$ = 2.24~mJ/m$^2$, anisotropy $K$ = 8.63$\times$10$^5$~J/m$^3$, saturation magnetization M$_s$ = 1.37$\times$10$^6$ A/m and the Gilbert damping $\alpha$= 0.3. We relax the micromagnetic state of a 100-nm-wide nano-disk from an initial uniform out-of-plane ferromagnetic state (FM). The obtained magnetization distributions at different external magnetic fields and temperatures are presented in Fig.~\ref{5}. One can see that without external field, the labyrinth domains with chiral N\'eel domain walls are established in the h-BN/Co nanodisk. More importantly, the labyrinth domains can be transformed into skyrmion states by applying external field as in many typical ferromagnet/heavy metal (FM/HM) heterostructures~\cite{Moreau-Luchaire2016,Soumyanarayanan2017,Boulle2016} with skyrmions. When the external field is in the range of about 200-250 mT, isolated skyrmions are induced in the nano-disk [Fig.~\ref{5}(a)] that can sustain temperatures up to above 100 K [Fig.~\ref{5}(b)]. These results suggest the h-BN/Co thin films can serve as a promising system to induce skyrmions avoiding heavy metals.

In conclusion,  we demonstrated the significant and robust behavior of PMA and DMI at interfaces comprising Co and 2D materials. By comparing h-BN/Co and Gr/Co interfaces, while the PMA behavior shows similarities for both cases, the DMI at Co/h-BN interface shows significantly larger values remaining robust over large Co thickness range unlike for Co/Gr interface. This is due to Rashba splitting in case of Co/h-BN found to be at least twice larger than that of Co/Gr. The physical mechanisms of such behavior are attributed to existence of competing dipoles at the Co/Gr interface compared to only one dipole at the opposite Co/h-BN interface which gives rise to a larger DMI. Moreover, based on found DMI and PMA behavior the possibility of skyrmions formation at h-BN(AC)/Co interface with the application of small external magnetic field and stable up to hundred Kelvin, is demonstrated. These findings demonstrate that 2D materials such as h-BN provide a viable alternative for heavy metals in the next-generation spintronic devices based on domain wall or skyrmoins.

\section{Acknowledgments}
This work was supported by the European Union's Horizon 2020 research and innovation Programme under grant agreement No. 785219 (Graphene Flagship).
\section{Methods}
First-principles calculations have been performed using the VASP (Vienna {\it Ab-initio} simulation package)~\cite{Kresse1993,Kresse1996,Kresse1996prb} with electron-core interactions described by the projector augmented wave method~\cite{Blochl1994}, and the exchange correlation energy calculated within the generalized gradient approximation of Perdew-Burke-Ernzerhof (PBE)~\cite{Perdew1996}. The cutoff energies for the plane wave basis set used to expand the Khon-Sham orbitals were chosen to be 520~eV for all calculations. The lattice constant of bulk Co(0001) hexagonal closed packed (hcp) structures is about 2.507~\AA{} that cannot fit perfectly on graphene and h-BN monolayers that have lattice constant of 2.46 and 2.504~\AA, respectively. In all calculations, the in-plane constant was fixed to be 2.507~\AA{}. The supercell comprised a varied number of Co layers (1-7 ML) coated by 1 ML of graphene or h-BN and followed by a vacuum region that is large enough to prevent interaction with the mirror image. For the surface magnetic anisotropy and DMI calculations a Monkhorst-Pack scheme was used for the $\Gamma$-centred 21$\times$21$\times$1  and 4$\times$16$\times$1 $K$-point mesh, respectively. Calculations were performed in three steps. First, the structural relaxations were performed until the forces become smaller than $0.001$~eV/\AA{} for determining the most stable interfacial geometries. Next, the Khon-Sham equation were solved, with no SOC, to find out the charge distribution of the system's ground state. Finally, SOC was included and the self-consistent total energy of the system was determined as function of the orientation of the magnetic moments. For DMI calculations, the orientation of the magnetic moments for each atom were controlled using the constrained method as implemented in VASP.
The structure of Co/h-BN films investigated is shown in Fig.~\ref{3} where a layer of graphene or h-BN covers the surface of hexagonal closed-packed (hcp) Co (0001)  film.  The interlayer distance between Co and the 2D materials was found to be 2.11\AA{} and 2.15\AA{} for graphene and h-BN, respectively. For Gr/Co interface, the most stable geometry configuration is found to be AC stacking. For h-BN/Co, the calculated ground state structure is found to be AB staking while the AC stacking is quasi stable with a small difference between the two configurations in the order of few meV/\AA. Both configurations denoted as h-BN(AC)/Co and h-BN(AB)/Co interfaces are considered in this study.

\bibliography{h-BN_Co_DMI_Skyrmions}

\end{document}